\documentclass[a4paper,10pt]{article}

\usepackage[latin1]{inputenc}
\usepackage{graphicx}

\def\FPU{Fermi--Pasta--Ulam\ }
\def\fraz#1#2{ {#1 \over #2} }

\title{On the definition of temperature in FPU systems.}
\author{A. Carati\footnote{ Universit\`a di Milano, Dipartimento di Matematica
Via Saldini 50,  20133 Milano (Italy)  E-mail: {\tt carati@mat.unimi.it} },
P. Cipriani\footnote{E-mail: {\tt cipriani@icra.it}} 
and L. Galgani\footnote{ Universit\`a di Milano,
Dipartimento di Matematica, Via Saldini 50,  20133 Milano (Italy)  
E-mail: {\tt galgani@mat.unimi.it} } }

\begin{document}

\maketitle


\centerline {ABSTRACT} 
\noindent  
\vskip 1truecm
It is usually assumed, in classical statistical mechanics, that the
temperature should coincide, apart from a suitable constant
factor, with the mean 
kinetic energy of the particles. We show that this is not the case for
\FPU systems, in conditions in which energy equipartition between the modes 
is not attained. We find that the temperature should be
rather identified with the mean value of the energy of the low frequency modes.

              \par
\noindent Running title: Temperature in FPU systems  


\section{Introduction.} The \FPU system consists of a chain of $N$ nonlinear
oscillators with certain given boundary conditions, tipically fixed ends. 
It is well known (see \cite{bocloi} and \cite{berchialla}) that, 
for energies below a certain threshold $E_{\rm c}$,
if the energy is initially given to a few low frequency modes,
equipartition of energy among the modes is eventually attained
only after an extremely long time, while at intermediate times a kind
of metaequilibrium state is attained, in which the energy is shared
essentially within a packet of low frequency modes.

An interesting and much discussed problem, is whether the specific
energy threshold $E_{\rm c}/N$ vanishes or not in the limit of an
infinite number of particles. Here we leave this problem aside: we
will suppose for example that the number $N$ of particles be fixed, so
that the threshold certainly exists.  We address instead the problem
raised by the fact that below the threshold one meets, as in the theory of
glasses, with time scales to thermal equilibrium which are very long,
even longer than any fixed observational time scale.  Does this lack
of thermalization have any consequences on the relevant
thermodynamical quantities? or even, is it possible to correctly
(i.e. uniquely) define the quantities of interest?  In other words, is
it still possible to have a thermodynamics below the threshold?

In the literature, the discussion is usually focused on the specific
heat, because from heuristic arguments it is suggested that to less
chaotic motions there correspond smaller specific heats, with
eventually zero specific heat for totally ordered motions (i.e. for
integrable systems).  Thus it is expected that by lowering the energy
below threshold the specific heat should diminish; such a property, in
turn, should be considered as a good indicator of the weakening of
chaos.

The papers \cite{vulpio} and \cite{tene}, which aim at evaluating the
specific heat of \FPU systems below threshold by numerical computations,   
reach two opposite
conclusions: in \cite{vulpio} the value of the specific heat remains
constant (as would follow from the equipartition principle) even below
the threshold, while in \cite{tene} the specific heat indeed begins to
fall down, below the threshold, and finally vanishes as temperature
goes to zero.  This striking difference is apparently due to the
different methods used in the two papers in estimating the specific
heat. Actually, in both papers the \FPU system is kept isolated (fixed
ends), so that a direct measurement of the specific heat is precluded
(because a direct measurement requires at least one heat bath).  The
specific heat is then estimated from the fluctuations of energy of a
subsystem through the well known relation between specific heat and
mean square deviation of energy, which holds in the canonical
ensemble.  The two papers differ in the choice of the subsytem: in the
paper \cite{vulpio}, one considers the energy fluctuations of a small
piece of the total chain, while in \cite{tene} one considers the
energy fluctuations of a small packet of nearby modes. As the energy
of each mode remains nearly constant below the threshold while the
energy of a piece of chains still present large fluctuations, this
indeed explains why the two papers reach opposite conclusions.  Now,
at most only one of the conclusions can be correct, if a right
conclusion does exist at all; indeed it is not clear whether the
specific heat can be defined in an unambiguous way below the threshold.

A different approach was proposed in the paper \cite{fpu}. In short,
the basic remark is the following one. As the above mentioned relation
between specific heat and energy fluctuations is obtained from the
equilibrium Gibbs ensemble, then its validity below the threshold is
in doubt just because, up to the considered times, the system has not
yet thermalized.  So, it is argued that in measuring the specific heat
one should revert to the conventional method which makes use of a heat
bath at a given temperature $T$ coupled to the \FPU system, with the
corresponding familiar calorimetric expression for the specific
heat. Namely, the energy exchange $\Delta Q$ between the bath and the
\FPU system is measured when the temperature is varied by $\Delta T$,
then the ratio $\Delta Q/\Delta T$ is computed, and (in principle) the
limit is taken for vanishing $\Delta T$. However, even with such a
method, one is still confronted with a delicate problem, because the
amount of exchanged energy $\Delta Q$ does depend on how much time one
has waited in making the measurement (this is the so--called waiting
time of the theory of glasses). The curve predicted by equipartition
is recovered for infinitely long waiting times, while for finite
waiting times the specific heat is expected to vanish at sufficiently
low temperatures.

The question is thus: does there exist a natural way to choose a
definite waiting time? Another problem then arises, due to the fact
that, in the ratio $\Delta Q/\Delta T$ defining the specific heat, one
should insert in the denominator the variation of temperature of the
\FPU system and not that of the bath. The question is then whether the
temperature of the \FPU system is the same as that of the heat bath.
The very fact that the quantity $\Delta Q$ depends on the waiting time
actually shows that this is not the case, just because the equality of
the two temperatures would imply $\Delta Q=0$.  On the other hand, if
one were able to identify the temperature of the \FPU system, then the
question of the waiting time would have a quick answer: one should
wait until the heat bath temperature and that of the \FPU system have
become equal, and only at that time should one measure the
corresponding energy exchange.  So the possibility of having available
well defined thermodynamic quantities on short time scales is based on
the possibility of providing a good notion of temperature for the FPU
system before complete equipartition be achieved.  From this point of
view, the identification usually made of the temperature of the \FPU
system with the mean kinetic energy of its particles is not the
correct one, because in such a case the temperatures of the two
systems (\FPU system and heat bath) would remain different for
extremely long times.

The identification of the mean kinetic energy with temperature is so
deeply rooted in our minds, that the existence of another quantity
playing that role seems hardly conceivable.  The aim of this paper is
to show instead that this is possible.

In Section~2 we give a preliminary discussion of the zeroth law for
states of metaequilibrium such as those of the \FPU system below
threshold, in Section~3 we describe the model we employ for measuring of
the temperature of the \FPU system through heat baths by numerical
computations, and the numerical results are given in Section~4.

\section{Zeroth law and  temperature in states of 
metaequilibrium.} One of the basic features of 
thermal equilibrium  is the so--called zeroth law, which essentially
amounts to the transitivity of the equilibrium.  From this follows
(see \cite{libro}) that for any system there exists a function of its
macroscopic (the so called \emph{empirical temperature}) which 
has the same value for bodies in equilibrium. 

However, it is not granted that, for a given macroscopic state, the
equilibrium is unique if some of the internal degrees of freedom are
dynamically frozen.  We are thinking typically of the case of
polyatomic molecules (see \cite{prl}) for which it is known that the
exchanges of energy between the center of mass and the internal degrees
of freedom are so slow that the number of  effective degrees of
freedom depends on the time of observation. This is actually the
general  situation that occurs in states of  metaequilibrium.

We thus address the problem whether it is possible to have a zeroth
principle, and so also an empirical temperature, in situations of
metaequilibrium, in
which the physical quantities are changing only on a
very long time scale.  So, if we put  our \FPU system in heat contact
with another body, and observe that at first there is a rapid relaxation
to a certain state, while a later evolution to a final equilibrium
would take place  on a time scale much longer
than our observational scale, we can think of our system as if it were
equilibrated. Obviously, one is not granted that in such a situation
the zeroth law, i.e. the transitivity of this metaequilibrium state,
holds. But, if this is the case,  an empirical temperature can  be
defined. In other terms, if we put  the \FPU system in contact  with a
thermometric body which, after a short transient, appears to have
reached a temperature $T$ (not evolving on our time scale),
and if later, after having put  the \FPU system in contact with another
body at the same temperature $T$, nothing seems to occur (i.e. there
is no exchange of energy in mean between the bodies), then the
metastable state reached does have the transitive property, and we are
authorized to assign to the \FPU system the temperature $T$ reached by
the thermometric body.

An equivalent arrangement, which we have actually implemented in our
numerical simulations to be described below, is the following one: the
\FPU system is put at the same time in contact both with a heat bath
and with a thermometric body (obviously, with no direct connection
between the two external bodies).  In such a situation, the transitive
property reduces to the property that in a short time the thermometer
attains the same temperature of the heath bath; the subsequent
evolution to a final equilibrium should take place later, at a much
smaller rate. Then, \emph{by definition}, the temperature of the \FPU
system in the metaequilibrium state is the one reached by the
thermometer after the short--time relaxation.

This definition can appear satisfactory from an operative point of
view. However, as it stands, it still lacks a clear connection with
the properties of the \FPU system itself. Indeed there remains the
problem of understanding, how the zeroth law can hold even if the
\FPU system did not yet thermalize. In this connection, we make
reference to a known phenomenon \cite{berchialla} concerning the
isolated \FPU system, namely the fact that, below threshold, for
initial excitations of the low frequency modes the energy turns out to
remain confined to modes below a certain critical mode $k_{{\rm cr}}$,
while the higher modes are not significantly involved in the dynamics.
Notice, moreover, that an analogous phenomenon, i.e. a dynamical
involvement restricted to the modes of sufficiently low frequency, is
know to occur also when a system is coupled to an external body, for
the case of polyatomic molecules (see \cite{landauteller}).  So, it is
known that, on a short time scale, the high frequencies modes (above
$k_{{\rm cr}}$) do not get dynamically involved, neither by the
internal nonlinarities nor by an interaction with external bodies. In
both cases a packet of low frequency modes is formed which are in
mutual equilibrium, and moreover are active in the process of
thermalization with external bodies.  Only after a much larger time
scale there follows a relaxation of the \FPU system to the true
equilibrium state.  Before, it appears as if there existed an
adiabatic partition (of a dynamical nature) between low and high
frequency modes.

If this is the correct picture, it is clear to what property of the
\FPU system should our definition of temperature correspond: namely,
to the mean energy of each of the low frequency modes (those below
$k_{{\rm cr}}$).  In the rest of the paper we will illustrate the
results of some numerical computations, which, in our opinion,
strongly support the fact that such a metaequilibrium is transitive,
and that the empirical temperature thus defined coincides with the
energy of the low frequency modes.

\begin{figure}[htbp]
 \begin{center} \includegraphics[width=10cm]{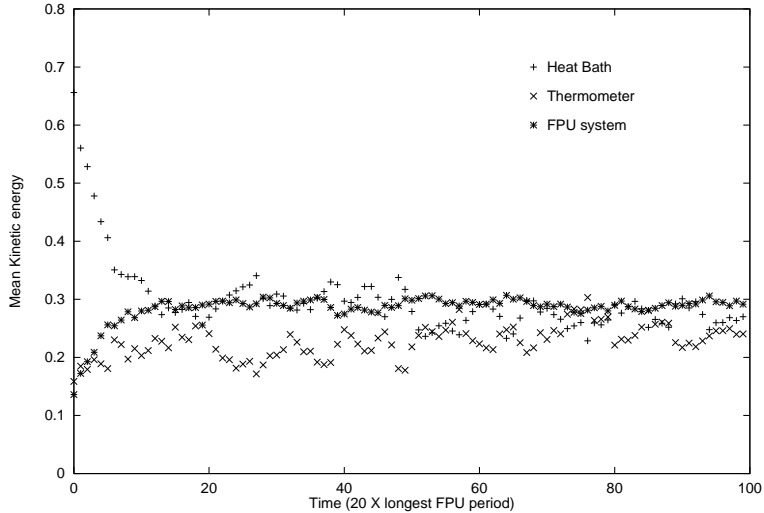}
 \end{center} \caption{Specific harmonic energy of the FPU system, and
 twice the specific energies of the gases, versus time, at high
 temperature.}  \label{Fig1}
\end{figure}

\section{The model.} As mentioned above, the model is constituted by a \FPU
system in contact with two bodies, acting, the one as a thermometer
and the other one as a heat bath. We make the simplest choice, in
which the two bodies are perfect gases. Each gas is modeled as a
system of point particles having no interactions among them, while
interacting with the \FPU system through some smooth force between
each of the gas particles and one of the edge \FPU particles.  In more
detail, concerning the \FPU system we denote as usual by $x_i$,
$i=1,{\ldots},N$, the distance of the $i$--th particle from its
equilibrium position, by $p_i$ its conjugate momentum, and consider the
familiar ``$\beta$--model'' Hamiltonian
\begin{equation}
H_{{\rm FPU}}=\sum_{i=1}^N \fraz {p_i^2}{2m} + \sum_{i=0}^N \fraz
      {\Omega^2}2 (x_{i+1}-x_i)^2 + \fraz \beta4 (x_{i+1}-x_i)^4 \ ,
\end{equation}
involving two positive parameters $\beta$ and $\Omega$, with fixed
boundary conditions $x_0=0$, $x_{N+1}=0$.  Now, to the `'left'' of the
\FPU chain we place a perfect gas which acts as a heat bath: denoting
by $y_i$, $\pi_i$ the $i$--th gas particle's position and momentum
respectively, we have $-L<y_i<x_1$, with $L>0$ playing the role of the
volume of the gas. The motion of each particle is thus free apart from
the fact that it suffers an elastic reflection as $y_i= -L$, and
that it moreover interacts with the first \FPU particle $x_1$ through a
short range potential, which we choose as
$$
V= V_0 \fraz{ e^{-(y_i-x_1)/l_0}}{ (y_i-x_1)/l_0} \ ,
$$
$l_0$ and $V_0$ denoting its range and strength respectively.  In
agreement with the bound given above, due to the singularity of the
potential at $y_i=x_1$, the solutions $y_i(t)$ of the equations of
motion cannot cross the point $x_1(t)$, i.e. for all times $t$ one has
$-L<y_i(t)<x_1(t)$. The Hamiltonian of the heat bath is thus
\begin{equation}
H_{{\rm B}}=\sum_{i=1}^N \fraz {\pi_i^2}{2m} + V_0 l_0 \fraz{
     e^{-(y_i-x_1)/l_0}}{ (y_i-x_1)} \ ,
\end{equation}
supplemented by the boundary condition that the particles are
reflected at $y_i=L$.

\begin{figure}[htbp]
 \begin{center}
 \includegraphics[width=10cm]{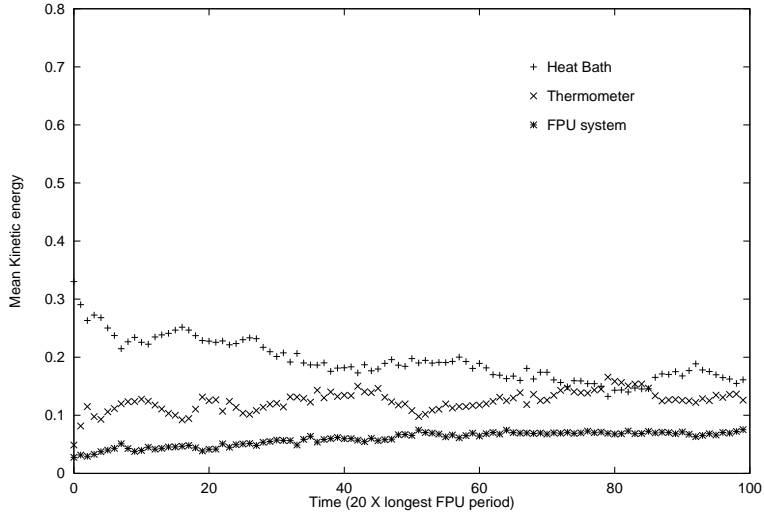}
 \end{center}
 \caption{Same as Figure~1, at low temperature.} 
 \label{Fig2}
\end{figure}

To the `'right'' of the \FPU chain we place the thermometer, which is
taken again as a perfect gas, with Hamiltonian
\begin{equation}
H_{{\rm T}}=\sum_{i=1}^N \fraz {\tilde \pi_i^2}{2m} + V_0 l_0\fraz{
     e^{-(\tilde y_i-x_N)/l_0}}{ (\tilde y_i-x_N)}\ ,
\end{equation}
(plus a reflection condition at $\tilde y_i =L$) where $\tilde y_i$,
$\tilde \pi_i$ are the positions and momenta of the gas particles
respectively; each of the particles interacts only with the last
particle $x_N$ of the \FPU system via the same potential as for the
heath bath.

\begin{figure}[htbp]
 \begin{center}
 \includegraphics[width=10cm]{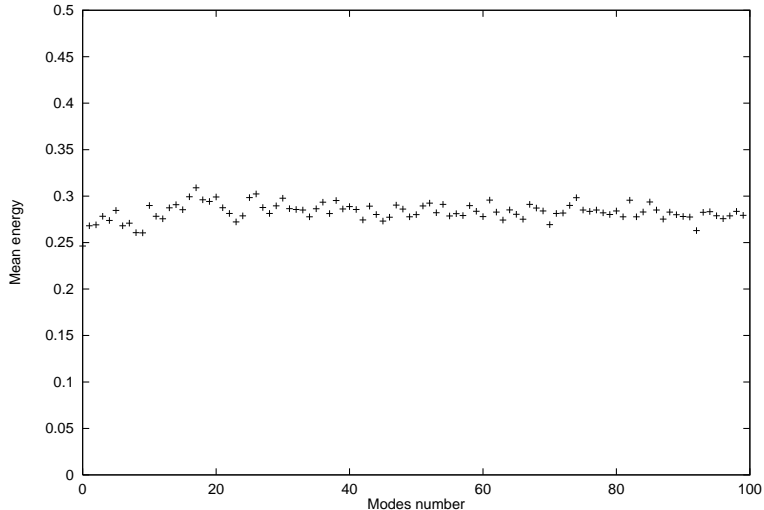}
 \end{center}
 \caption{Energy spectrum of the FPU system, at high temperature.}
 \label{Fig3}
\end{figure}

In our simulations we chose an equal number of particles for the three
systems, while in principle the number of bath particles should be
larger than that of the \FPU system, and this in turn should be larger
than that of the thermometer.  Our choice is dictated only by the
computational power available: we cannot take the total number of
particles too large, but at the same time the number of particles in
each system cannot be too small if a good statistics has to be
insured. Taking the same number $N=100$ of particles for each of the
three subsystems, seemed to us a good compromise.

We took as units of mass, length and energy the values $m$, $l_0$ and
$V_0$, which were thus put equal to one in our computations. The
values of the parameters $\Omega$ and $\beta$ were set equal to
$\Omega=400$ and $\beta=3742$ respectively.  Such strange values come
from the following consideration: the intermolecular interaction in a
crystal is well represented by a Lennard--Jones potential, whose
relevant parameters (the range and the strength) are of order one with
our choice of units. On the other hand the \FPU potential should just
be a Taylor expansion of the Lennard--Jones potential around the
equilibrium position.  Performing such a Taylor expansion and putting
the parameters equal to one, the indicated values for $\Omega$ and
$\beta$ are found.

Finally in our numerical simulations we took $L=25$; this in order to
ensure a sufficient total number of collisions (of the order $10^6$ in
our actual integrations), while at the same time letting the gas
particles be free for a large part of their paths.

\section{Numerical results.}
The integration step was taken equal to a twentieth of the shortest
period $\tau_{{\rm f}}= \pi/\Omega$ of the \FPU chain, and the
numerical solutions were computed up to times of order
$2\cdot10^7\tau_{{\rm f}}$.

The numerical experiments were performed in the following way. For the
bath we chose a temperature $T_1$ and took random initial conditions
extracted from a Maxwellian at the chosen temperature $T_1$ (we also
checked that the value of the mean kinetic energy should not deviate
too much from the expected one, in order to avoid too large
fluctuations); for the \FPU system we chose initial data at
equipartition with a temperature $T_1/10$ and random phases; and
finally for the thermometer we chose initial data in the same way as
for the heath bath, but at a temperature $T_1/10$.  We let the system
evolve for a time $10^4\tau_{{\rm f}}$, and then began to compute the
time averages of the kinetic energies of the gases and of the harmonic
energy of the \FPU system.  The results of the computations for two
representative cases are shown in Figures~\ref{Fig1} and \ref{Fig2},
where we report, versus time, the temperatures (i.e. twice the kinetic
energies per particle) of the gases and the harmonic energy per
particle of the \FPU system (actually, time averages of such
quantities are reported).  Let us recall that such three quantities
should be equal according to the equipartition principle, i.e. for
sufficiently long times.

\begin{figure}[htbp]
 \begin{center}
 \includegraphics[width=10cm]{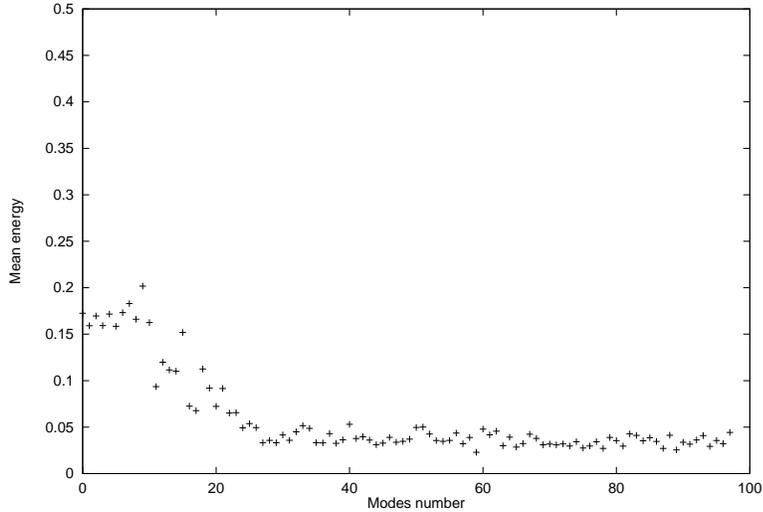}
 \end{center}
 \caption{Same as Figure~3, at low temperature.}
 \label{Fig4}
\end{figure}

In Figure~\ref{Fig1}, we started with a temperature $T_1=1$. One sees
that, after a time of order $10^5\tau_{{\rm f}}$, the
temperatures of two gases and the specific harmonic energy of the \FPU
system have become essentially equal, although still presenting
significant fluctuations.  This case should correspond to a situation
of thermal equilibrium.

Figure~\ref{Fig2} refers instead to the choice of $T_1=0.4$. One sees
that the heat bath and the thermometer still reach the same
temperature (although after a time almost one order of magnitude
larger than before), while the specific harmonic energy per particle
of the \FPU system remains well below the common temperature of the
two gases, up to the observation time.  Actually the curve is so flat
that the \FPU system can be expected to possibly reach the equilibrium
only on a totally different time scale. The global system seems indeed
to be in a situation of metaequilibrium.

It appears, however, that the zeroth law can still be valid, and that
the temperature ``measured'' by the thermometer is a good empirical
one, because the temperatures of the two gases have become equal. To
understand to which quantity of the \FPU system does this measured
temperature correspond, in Figures~\ref{Fig3} and \ref{Fig4} we report
the spectra (time--averaged energies of the modes versus mode number)
of the \FPU system at the end of the two runs.  Figure~\ref{Fig3}
refers to the case of complete thermalization, and correspondingly a
complete equipartition among the modes is found, as expected.  More
interesting is Figure~\ref{Fig4}: here equipartition obtains only
among modes of sufficiently low frequency, say below $k_{{\rm
cr}}=10$, while the energy starts decreasing for larger values of $k$,
going down, say for $k>25$, to the initial equipartition value $0.04$. 
It does not appear as a surprise to observe that the mean energy of
the low frequency modes essentially agrees with the common
temperature of the two gases. This seems to indicate that, for our
metastable state, the ``good'' definition of temperature of the \FPU
system is the mean energy of the (sufficiently) low frequency modes.

\section{Conclusions.} In conclusion, 
we hope to have shown, through our numerical study of a \FPU system in
contact with two gases, that there are cases of metastable equilibrium
for which a notion of temperature can be defined. However, at variance
with the familiar case of equilibrium, such a temperature does not
coincide with the ``canonical'' one, namely twice the mean kinetic
energy per particle.

As a further comment, we would like to add that the metaequilibrium
states met in \FPU systems present characteristics which are somehow
opposite to those of glasses. Indeed, in the latter case the lack of
thermalization is ascribed to the low frequency modes and
correspondingly the thermometer measures the mean energy of the high
frequency modes, which are the one being in mutual equipartition.

Finally, we would like to mention that the possibility of having a
thermodynamics for situations of metaequilibrium, typically
involving the presence of adiabatic invariants, was amply discussed in
the second part of a very interesting paper of Poincar\'e \cite{poin},
which appears to have been almost completely forgotten.  In fact we
became aware of such work only after completing the present work,
through a conference of V. Kozlov \cite{kozlov}. In fact, V. Kozlov was
addressing only the problem dealt with in the first part of the paper of
Poincar\'e, namely how it occurs that the fast variables of an
integrable hamiltonian system approach equilibrium, notwithstanding
the reversibility and the return property of the system.  In the
second part of his paper, Poincar\'e was instead considering a
situation in which one has at first a quick relaxation to a
``provisional equilibrium'' while a ``definitive equilibrium'' would
be attained after a much larger time, i.e. one is concerned, in his
very words, with ``very long times of first order'' and ``very long
times of second order'', which is a situation analogous to the one
discussed in the present paper.


\end{document}